\documentclass[conference]{IEEEtran}
\IEEEoverridecommandlockouts
\usepackage{cite}
\usepackage{amsmath,amssymb,amsfonts}
\usepackage{algorithm}
\usepackage{algpseudocode}
\usepackage{graphicx}
\usepackage{psfrag}

\graphicspath{{./images/}}
\DeclareGraphicsExtensions{.pdf,.jpeg,.png}
\usepackage{textcomp}
\usepackage{xcolor}
\def\BibTeX{{\rm B\kern-.05em{\sc i\kern-.025em b}\kern-.08em
    T\kern-.1667em\lower.7ex\hbox{E}\kern-.125emX}}
\newcommand{\tildebf}[1]{\Tilde{\mathbf{#1}}}

\newcounter{tempEquationCounter}
\newcounter{thisEquationNumber}

\begin{document}

\title{Low-Complexity Equalization and Detection for OTFS-NOMA \\
}

\author{\IEEEauthorblockN{Stephen McWade\IEEEauthorrefmark{1}, Mark F. Flanagan\IEEEauthorrefmark{2},
 and 
Arman Farhang\IEEEauthorrefmark{1}}

\IEEEauthorblockA{\IEEEauthorrefmark{1}Department of Electronic and Electrical Engineering, Trinity College Dublin, Ireland \\}

\IEEEauthorblockA{\IEEEauthorrefmark{2}School of Electrical and Electronic Engineering, University College Dublin, Ireland \\
Email: \{smcwade, arman.farhang\}@tcd.ie, mark.flanagan@ieee.org}

\thanks{
This publication has emanated from research conducted with the financial support of Science Foundation Ireland (SFI) under Grant number 17/RC-PhD/3479, Grant number 17/US/3445 and Grant number 19/FFP/7005(T).
}
}

\maketitle

\begin{abstract}
    Orthogonal time frequency space (OTFS) modulation has recently emerged as a potential 6G candidate waveform which provides improved performance in high-mobility scenarios.  In this paper we investigate the combination of OTFS with non-orthogonal multiple access (NOMA). Existing equalization and detection methods for OTFS-NOMA, such as minimum-mean-squared error with successive interference cancellation (MMSE-SIC), suffer from poor performance. Additionally, existing iterative methods for single-user OTFS based on low-complexity iterative least-squares solvers are not directly applicable to the NOMA scenario due to the presence of multi-user interference (MUI). Motivated by this, in this paper we propose a low-complexity method for equalization and detection for OTFS-NOMA. Our proposed method uses an iterative process where in each iteration, an equalizer based on the least-squares with QR factorization (LSQR) algorithm is followed by a novel reliability zone (RZ) detection scheme which estimates the reliable symbols of the users and then uses interference cancellation to remove MUI. We present numerical results which demonstrate the superiority of our proposed method, in terms of symbol error rate (SER), to the existing MMSE-SIC benchmark scheme. Additionally, we present results which illustrate that a judicious choice of RZ thresholds is important for optimizing the performance of the proposed algorithm.
\end{abstract}


\section{Introduction}
The sixth generation (6G) of mobile networks is expected to support communications in high-mobility environments such as high-speed rail, vehicle-to-everything (V2X) and unmanned aerial vehicle (UAV) communications \cite{Tataria_6G}. Orthogonal frequency division multiplexing (OFDM) has been the waveform utilized in the 4th and 5th generation of wireless networks. However, it is well-known that in high-mobility scenarios, OFDM performs poorly due to the Doppler effect \cite{Wei_OTFS}. In recent years, a new waveform called orthogonal time frequency space (OTFS) has been proposed to address this drawback of OFDM in time-varying channels. In contrast to OFDM, which transmits data symbols in the time-frequency domain, OTFS places the data symbols in the delay-Doppler domain \cite{Hadani_OTFS}. OTFS then uses a transformation to spread each information symbol over the time-frequency plane. This means that the symbols are all equally affected by the time and frequency selectivity of the channel which converts the time-varying channel to a time-invariant one in the delay-Doppler domain.

 A number of OTFS equalization and detection schemes have been proposed in the literature in recent years. The majority of methods can be categorized into either low-complexity linear equalizers \cite{Tiwar_OTFS_LMMSE,Surabhi_OTFS_EQ,ZOU_OTFS_EQ} or non-linear message-passing-based equalizers \cite{Rav_INT_canc_MP, Surabhi_MP_mmWAVE, MIMO_OTFS_MP}. However, these methods assume a scattering environment in which the channel impulse response is sparse in the delay-Doppler domain. Under more realistic channel conditions, the low-complexity linear schemes are no longer applicable and message-passing-based detectors become prohibitively complex due to the large number of scatterers \cite{Qu_OTFS_detection}. An alternative approach was proposed by the authors of \cite{Qu_OTFS_detection} which utilized a least-squares minimum residual (LSMR) based
channel equalizer and a reliability-based
dynamic detector to estimate the transmitted data symbols. However, the system model in \cite{Qu_OTFS_detection} only considers a single-user scenario.

For a multi-user OTFS system, the multiple access (MA) technique utilized is an important consideration. The methods proposed thus far in the literature can be broadly categorized as orthogonal multiple access (OMA)  or non-orthogonal multiple access (NOMA). In OTFS-OMA, users are multiplexed either in the delay or the Doppler domain and only one user can occupy a given resource block \cite{Chong_OTFS_Uplink_Rate}. However, the users suffer from multi-user interference (MUI) due to the Doppler spread. The resulting MUI can be mitigated by inserting guard bands between the users, as was done in \cite{patent_Hadani}. However, this leads to a spectral efficiency (SE) loss \cite{surabhi2019multiple}. 

An alternative approach is OTFS-NOMA, where the users are allowed to occupy the same resource block and are multiplexed in either the power domain or the code domain. A multi-user detection (MUD) scheme such as successive interference cancellation (SIC) is then used to detect the user symbols \cite{DaiSurveyNOMA}. NOMA is a well-known technique which can provide improved SE compared to OMA. A number of OTFS-NOMA schemes have been proposed in the literature in recent years which use either power-domain \cite{Poor_OTFS_NOMA, Chatt_OTFS_NOMA} or code-domain \cite{OTFS_SCMA, OTFS_code_noma} multiplexing. In this paper, we focus on power-domain OTFS-NOMA. 

With regard to the existing work on power-domain OTFS-NOMA, the authors of \cite{Poor_OTFS_NOMA} considered a single high-mobility OTFS user multiplexed with multiple low-mobility OFDM users. However, this system model is restricted to a single OTFS user and hence cannot accommodate multiple high-mobility users. The authors of \cite{Chatt_OTFS_NOMA} addressed this issue and proposed an OTFS-NOMA scheme which utilizes a rectangular pulse shape where multiple users overlap in the delay-Doppler domain and are multiplexed in the power domain. However, the system proposed in \cite{Chatt_OTFS_NOMA} used minimum-mean-squared-error (MMSE) equalization in combination with SIC for equalization and detection. The problem with this scheme is that direct implementation of MMSE for equalization is prohibitively computationally complex and thus impractical for real-world scenarios. 

 As of yet, there is no low-complexity equalization and detection method for power-domain OTFS-NOMA. In particular, the low-complexity equalization and detection method of \cite{Qu_OTFS_detection} cannot be directly applied to a NOMA scenario due to the presence of MUI.
This paper addresses these gaps in the literature with the following contributions: (i) we propose a low-complexity receiver for a 2-user downlink OTFS-NOMA system in which both users deploy OTFS and are multiplexed in the power domain; (ii) we propose an iterative detector for OTFS-NOMA which uses the least-squares with QR factorization (LSQR) algorithm to estimate the transmit symbols and a reliability zone (RZ) detection scheme on the symbols of each user; (iii) the proposed method uses a novel iterative process in which it detects symbols from each user within each iteration and uses interference cancellation to remove MUI as well as inter-Doppler interference (IDI) and inter-symbol interference (ISI). We present numerical results which show that the proposed method provides symbol error rate (SER) performance gains of up to 6~dB over MMSE equalization with SIC. Additionally, we present numerical results which demonstrate that the choice of thresholds within the RZ detector is important for obtaining high-quality detection.

 The rest of this paper is organized as follows. Section~II describes the system model for a 2-user OTFS-NOMA system. In Section~III, the proposed equalization and detection algorithm is presented. Section~IV presents numerical results. Finally, Section~V concludes the paper. 

 \subsubsection*{Notations} Superscripts ${(\cdot)^{\rm{T}}}$ and ${(\cdot)^{\rm{H}}}$ denote transpose and Hermitian transpose, respectively. Bold lower-case characters are used to denote vectors and bold upper-case characters are used to denote matrices. $x[n]$ denotes the $n$-th element of the vector $\mathbf{x}$. The function $\rm{vec}\{\mathbf{X}\}$ vectorizes the matrix $\mathbf{X}$ by stacking its columns to form a single column vector, and $\otimes$ represents the Kronecker product. The $p\times{p}$ identity matrix and $p \times q$ all-zero matrix are  denoted by $\mathbf{I}_p$ and $\mathbf{0}_{p\times{q}}$, respectively. Finally, $j = \sqrt{-1}$ represents the imaginary unit.
 
\section{System Model}
For ease of exposition, in the following sections we will describe the system model and the proposed detector for the case of a 2-user OTFS-NOMA system; however, note that the proposed method is generalizable to any number of users. We consider a downlink OTFS-NOMA system where both users occupy the same delay-Doppler domain resources and are multiplexed in the power domain. For User $i \in \{ 1,2 \}$, let the $M \times N$ matrix $\mathbf{X}_i$ contain the $MN$ quadrature amplitude modulation (QAM) data symbols placed in the delay-Doppler domain. The elements of $\mathbf{X}_i$ are assumed to be independent and identically distributed (i.i.d.) complex random variables. Additionally, a normalized (unit-energy) square QAM constellation is assumed for each user.

The delay-time domain transmit signal of User $i$ can be expressed as $   \mathbf{S}_i = {\mathbf{A}_{\mathrm{cp}}}\mathbf{X}_i\mathbf{F}_{N}^{\rm{H}},$ where $\mathbf{F}_N$ is the $N$-point unitary discrete Fourier transform (DFT) matrix in which the $(l,k)$ element is $\frac{1}{\sqrt{N}}e^{-j\frac{2\pi}{N}lk}$ and ${\mathbf{A}_{\mathrm{cp}}} = \left[\mathbf{J}_{\rm{cp}}, \mathbf{I}_{N}\right]$ is the CP addition matrix (here $\mathbf{J}_{\rm{cp}}$ is composed of the last $N_{\mathrm{cp}}$ rows of $\mathbf{I}_{N}$).
After parallel to serial conversion, the time-domain symbols for User $i$ can be written as
\begin{equation}
    \mathbf{s}_i = \mathrm{vec}(\mathbf{S}_i) =  (\mathbf{F}_{N}^{\mathrm{H}}\otimes{\mathbf{A}_{\mathrm{cp}}})\mathbf{x}_i.\label{eq:4}
\end{equation}

The users are then multiplexed in the power domain and their signals are superimposed before transmission. The superimposed transmit signal is given by $\mathbf{s} = \sqrt{{p}_1}\mathbf{s}_1 + \sqrt{{p}_2}\mathbf{s}_2,$
where ${p}_i$ is the power allocation coefficient for User $i$ and $p_1+p_2=1$. Additionally, we assume that User 1 has a lower average received signal-to-noise ratio (SNR) and therefore is allocated a larger proportion of the total power available, i.e., $p_1>p_2$ \cite{Chatt_OTFS_NOMA}. After digital to analog conversion, the continuous-time signal $s(t)$ is transmitted through the linear time-varying (LTV) channel. The received signal at the receiver for User $i \in \{ 1,2 \}$ can be written as
\begin{equation}
        r_i(t) = \int \int h_i(\tau,\nu)s(t-\tau)e^{j2\pi\nu(t-\tau)}d\tau d\nu + \omega_i(t),\label{eq:6}
\end{equation}
where $$h_i(\tau,\nu) = \sum_{p=0}^{P_i-1}h_{i,p}\delta(\tau - \tau_{i,p})\delta(\nu - \nu_{i,p}),$$ is the delay-Doppler channel impulse response (CIR) for User $i$, which consists of $P_i$ channel paths, and $\omega_i(t)$ is the complex AWGN with variance $\sigma_{i}^2$. The parameters $h_{i,p}$, $\tau_{i,p}$ and $\nu_{i,p}$ represent the channel gain, delay and Doppler shift, respectively, associated with path $p$ of User $i$'s channel. The received signal is then sampled with sampling period $T_{\rm{s}}$ and the discrete received signal samples can be expressed as 
 \begin{equation}
        r_i[n] =  \sum_{l=0}^{L-1}h_i[n,l]s[n-l] + \omega_i[n], \label{eq:7}
\end{equation}
where $h_i[n,l]$ is the CIR for User $i$'s channel at time instant $n$ and delay $l$. The discrete-time received signal for User $i$ can be written in matrix form as $\mathbf{r}_i = \mathbf{H}_i\mathbf{s} + \boldsymbol{\omega}_i,$
where $\boldsymbol{\omega}_i$ is the $MN \times 1$ complex AWGN vector and $\mathbf{H}_i$ is the $MN \times MN$ time-domain channel matrix of User $i$ constructed from the CIRs. The received signal is then demodulated and converted back to the delay-Doppler domain by performing an $N$-point DFT operation across the time-domain samples. Thus, the received signal is given by $\mathbf{y}_i = (\mathbf{F}_{N}\otimes\mathbf{R}_{\mathrm{cp}})\mathbf{r}_i.$
This can alternatively be written as $\mathbf{y}_i = \mathbf{G}_i\mathbf{x}_{\mathrm{sup}} + \mathbf{w}_i,$
where $\mathbf{G}_i=(\mathbf{F}_{N}\otimes\mathbf{R}_{\mathrm{cp}})\mathbf{H}_i(\mathbf{F}_{N}^{\mathrm{H}}\otimes\mathbf{A}_{\mathrm{cp}})$ is the effective channel matrix, $\mathbf{x}_{\mathrm{sup}}=\sqrt{p_1}\mathbf{x}_1 + \sqrt{p_2}\mathbf{x}_2$ is the superimposed delay-Doppler symbol vector and $\mathbf{w}_i=(\mathbf{F}_{N}\otimes\mathbf{R}_{\mathrm{cp}})\boldsymbol{\omega}_i$ is the noise vector.

\section{Proposed Equalization and Detection Technique}
The proposed method is inspired by the method proposed in \cite{Qu_OTFS_detection} for single-user OTFS which utilized an iterative LSMR based method with RZ detection and interference cancellation. However, if the method in \cite{Qu_OTFS_detection} is applied directly to OTFS-NOMA with SIC to detect the signals of User 1 and User 2, we can expect poor performance due to the MUI present in the system. Therefore, in the proposed method, we perform SIC at a symbol level rather than a packet level. This allows for the decoding of symbols from both users as soon as they become reliable and also allows for the incorporation of MUI cancellation to improve the detection performance. The proposed algorithm uses an iterative process in which  the LSQR algorithm is used to equalize the channel and an RZ detector is used to detect the reliable symbols of both User 1  \textit{and} User 2 within each iteration. Interference cancellation is then used to remove ISI, IDI \textit{and} MUI from the undetected symbols of both users, which improves the detection quality in subsequent iterations. In the next subsection, we give a brief overview of the LSQR algorithm.

\subsection{LSQR algorithm}
The proposed method uses the LSQR algorithm to equalize the channel and obtain an estimate of the superimposed transmit symbol. LSQR is a well-known iterative algorithm for solving linear problems of the form $\mathbf{y} = \mathbf{G}\mathbf{x} + \mathbf{w}$  \cite{LSQR}. At iteration $u$, LSQR constructs the vector $\mathbf{x}_u$ in the Krylov subspace 
\begin{equation*}
    \begin{split}
        \mathcal{K}(\mathbf{G}^{\mathrm{H}}\mathbf{G},\mathbf{G}^{\mathrm{H}}\mathbf{y}, u) = \mathrm{span}\{ & \mathbf{G}^{\mathrm{H}}\mathbf{y}, (\mathbf{G}^{\mathrm{H}}\mathbf{G})\mathbf{G}^{\mathrm{H}}\mathbf{y}, \ \dots \ , \\ & (\mathbf{G}^{\mathrm{H}}\mathbf{G})^{u-1}\mathbf{G}^{\mathrm{H}}\mathbf{y} \}
    \end{split}
\end{equation*}
which minimizes the norm of the residual, $||\mathbf{y} -\mathbf{G}\mathbf{x}_k ||$. LSQR can also be regularized by including the noise variance per dimension $\sigma^2$ as a damping parameter. After several iterations, LSQR provides performance similar to MMSE but with lower complexity \cite{Hrycak_LSQR_GMRES}. At each iteration, the LSQR algorithm  uses Golub-Kahan bidiagonalization and QR decomposition to obtain the estimate $\mathbf{x}_u$ \cite{Hrycak_LSQR_GMRES}. The authors of \cite{LSQR} proposed a simple recursive method for updating this estimate within each iteration. The iterative process continues until either the norm of the residual reaches a pre-determined tolerance, $\epsilon$, or the maximum number of iterations $U$ is reached. After LSQR is used to obtain the estimate of the transmitted signal, the proposed algorithm uses an RZ detector to obtain the reliable symbols of each user. In the next subsection, we describe the RZ detector.

\subsection{Reliability zone detector}

\begin{figure}[t]
    \centering
    \includegraphics[width = \columnwidth]{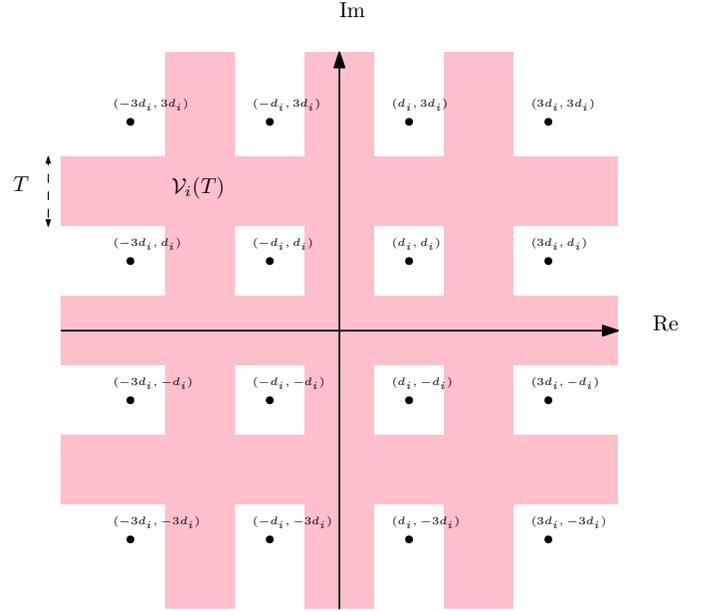}
    \caption{Illustration of the unreliable zone $\mathcal{V}_i(T)$ in  the case where User i employs a 16-QAM constellation.}
    \label{fig:RZ}
\end{figure}

Here, we first introduce some relevant notation. Each User $i \in \{1,2 \}$ uses $A_i$-ary QAM modulation, where the QAM symbol constellation is defined as 
\begin{equation*}
    \begin{split}
        \mathcal{A}_i =  \{&u + vj \ : \  u,v \in \{(2a-1)d_i : 
        \\& a \in \{-\sqrt{A_i}/2 +1 , \ \dots \ , \sqrt{A_i}/2 \}\}\},
    \end{split}
\end{equation*}
where $d_i$ is half the distance between adjacent QAM constellation symbols of User $i$. Next, we define the \textit{unreliable} zone with respect to this QAM constellation as
\begin{equation}
    \mathcal{V}_i(T) =  \{u + vj \ | u,v \in \mathcal{U}_i(T) \}, \label{eq:12}
\end{equation}
where
\begin{equation}
    \mathcal{U}_i(T)  =  \bigcup_{a=-\sqrt{A_i}/2 +1}^{\sqrt{A_i}/2 -1} \mathcal{U}_{i,a}(T) .
\end{equation}
and $$\mathcal{U}_{i,a}(T)  = \{u \ | \ 2ad_i-T/2 < u < 2ad_i + T/2 \},$$ where $T$ is a pre-defined threshold which determines the size of the unreliable zone. To demonstrate, the shaded areas in Fig. \ref{fig:RZ} shows an illustration of the unreliable zone $\mathcal{V}_i(T)$ for a 16-QAM constellation.

In the detection process, decisions are made in a symbol-by-symbol manner. If a symbol $x_i[n]$ is outside $\mathcal{V}_i(T_i)$, then it is deemed reliable and can be quantized to the nearest symbol in $\mathcal{A}_i$; the resulting symbol is denoted by $x_{i,q}[n] = Q_i(x[n])$. If $x_i[n]$ is inside $\mathcal{V}_i(T_i)$ then it is deemed unreliable and no quantization takes place. The detected reliable symbols can then be used for interference cancellation.

\subsection{Proposed equalization and detection algorithm}

In this subsection, we describe the proposed algorithm for equalization and detection of the OTFS-NOMA signal at User $i$'s receiver -  this is detailed in Algorithm \ref{PA1}. Each iteration begins on line 5 of Algorithm 1, where the LSQR algorithm is used to equalize the channel and obtain a new estimate of the superimposed transmitted symbol vector via 
\begin{equation}
    \tildebf{x}_{\mathrm{sup}}=\mathrm{LSQR}(\mathbf{G}_i, \mathbf{y}^{(k)}).
\end{equation}
\vspace{-2ex}
\begin{algorithm}
\caption{Proposed Algorithm for symbol detection at User~$i$ receiver}\label{PA1}
\begin{algorithmic}[1]
\State \textbf{Input}: User index $i$, Channel matrix $\mathbf{G}_i$, received symbol vector $\mathbf{y}_i$, power allocation fractions $p_1$ and $p_2$, starting thresholds $T_1^{(1)}$ and $T_2^{(1)}$
\State \textbf{Initialize}: $\mathbf{y}^{(1)}=\mathbf{y}_i$, $\mathbf{\hat{x}}_1= \mathbf{\hat{x}}_2 = \tildebf{x}_{1,\mathrm{q}}= \tildebf{x}_{2,\mathrm{q}}= \mathbf{0}_{MN\times1}$ 
\State Define $\mathcal{N}=\{0,\dots,MN-1\}$, $\mathcal{N}_1=\mathcal{N}_2=\mathcal{N}$, $\mathcal{D}_1= \emptyset$
\For{$k$ = 1 to $K$}
\State $\tildebf{x}_{\mathrm{sup}}=\mathrm{LSQR}(\mathbf{G}_i, \mathbf{y}^{(k)})$

\State $\tildebf{x}_{1} = ((\tildebf{x}_{\mathrm{sup}}[m])_{m\in\mathcal{N}_1})/\sqrt{p_1}$
\State $\tildebf{x}_{2}=((\tildebf{x}_{\mathrm{sup}}[m])_{m\in \{\mathcal{N}_2 \cup  \mathcal{D}_1\}})/\sqrt{p_2}$

\State Update users' reliable symbol index sets via $$\mathcal{R}_j = \{n \in \mathcal{N}_j :  \tildebf{x}_{j}[n] \notin \mathcal{V}_j(T_j^{(k)}) \}, \forall j \in \{1,2\}$$

\State Quantize reliable symbols: $$\tildebf{x}_{j,\mathrm{q}}[r] = Q_j(\tildebf{x}_{j}[r]), \ \forall r \in \mathcal{R}_j, \forall j \in \{1,2\}$$


\State Remove interference:
$$\mathbf{y}^{(k+1)}=\mathbf{y}^{(k)}-\mathbf{G}_i(\sqrt{p_1}\tildebf{x}_{1,\mathrm{q}} + \sqrt{p_2}\tildebf{x}_{2,\mathrm{q}})$$

\State Store detected symbols in output vectors:
$$\mathbf{\hat{x}}_j = \tildebf{x}_{j,\mathrm{q}}[r], \ \forall r \in \mathcal{R}_j, \forall j \in \{1,2\}$$

\State Reset: $\tildebf{x}_{1,\mathrm{q}}=\mathbf{0}$ and $\tildebf{x}_{2,\mathrm{q}}=\mathbf{0}$

\State Update: 
$\mathcal{N}_1=\{n \in \mathcal{N}: {\hat{x}}_1[n]=0\}$, $\mathcal{N}_2=\{n \in \mathcal{N}: {\hat{x}}_2[n]=0\}$, $\mathcal{D}_1=\{n \in \mathcal{N}: n \notin \mathcal{N}_1\}$

\State Update thresholds: $T_1^{(k+1)} = T_1^{(1)}(1-\frac{k+1}{K-1})$, and $T_2^{(k+1)} = T_2^{(1)}(1-\frac{k+1}{K})$

\State \textbf{if} $\mathcal{N}_i=\emptyset$, \textbf{break}

\EndFor
\State \textbf{Output}: $\mathbf{\hat{x}}_i$ 

\end{algorithmic}
\end{algorithm}

In lines 6 and 7, two sub-vectors are formed from $\tildebf{x}_{\mathrm{sup}}$. 
The vector $\tildebf{x}_{1}$ contains the elements of $\tildebf{x}_{\mathrm{sup}}$ whose indices are in $\mathcal{N}_1$, which is the set of undetected User 1 symbols. Since the RZ detector can only make decisions on User 2 symbols once the corresponding User 1 symbols have been detected on a previous iteration, the vector $\tildebf{x}_{2}$ contains the elements of $\tildebf{x}_{\mathrm{sup}}$ whose indices are in $\mathcal{N}_2$, the set of undetected User 2 symbols, \textit{and} $\mathcal{D}_1$, the set of detected User 1 symbols. In line 8, the RZ detector identifies the reliable symbols of both users, i.e., those on which we will make a decision in this iteration.

In line 9, the reliable symbols are quantized to the nearest QAM symbol and are stored in the empty vectors $\tildebf{x}_{1,\mathrm{q}}$ and $\tildebf{x}_{2,\mathrm{q}}$. In line 10, the quantized reliable symbols are used to remove interference from the received signal vector via
\begin{equation}
    \mathbf{y}^{(k+1)}=\mathbf{y}^{(k)}-\mathbf{G}_i(\sqrt{p_1}\tildebf{x}_{1,\mathrm{q}} + \sqrt{p_2}\tildebf{x}_{2,\mathrm{q}}).
\end{equation}
The quantized symbols are also stored in the estimated symbol vectors $\hat{\mathbf{x}}_1$ and $\hat{\mathbf{x}}_2$ (line 11). After canceling the interference from the detected symbols of both users, the algorithm updates the sets  $\mathcal{N}_1$ and $\mathcal{N}_2$ of undetected symbols, and the set  $\mathcal{D}_1$ of detected User 1 symbols based on the state of the output vectors $\mathbf{\hat{x}}_1$ and $\mathbf{\hat{x}}_2$. The thresholds for the RZ detector for each user  are then reduced geometrically within each iteration in line 14. This ensures that the algorithm converges within $K$ iterations and that all User $i$ symbols are detected. Since this is at User $i$'s receiver, the algorithm stops when all of the User $i$ symbols are detected, i.e., User 1 will detect all of its symbols before it detects all the User 2 symbols and can therefore stop once $\mathbf{\hat{x}}_1$ has no entries equal to zero.

\subsection{Computational complexity}
In this subsection, the computational complexity of the proposed method is compared to that of the MMSE-SIC benchmark, in terms of the number of complex multiplications. Direct implementation of MMSE equalization involves the inversion of an $MN \times MN$ matrix and hence has a computational complexity of $\mathcal{O}(M^3N^3)$. Each iteration of the LSQR algorithm has a computational complexity of $\mathcal{O}(MN\log_2(MN))$ \cite{Hrycak_LSQR_GMRES}. In the worst-case scenario Algorithm 1 performs LSQR $K$ times, each with $U$ LSQR iterations. Therefore, the computational complexity of Algorithm 1 is $\mathcal{O}(UKMN\log_2(MN))$. In practice, the typical values of $K$ and $U$ are in the order of tens and the typical value of $MN$ is in the order of thousands. Thus, $UK << M^2N^2$ and our method can achieve orders of magnitude computational complexity improvement over MMSE-SIC for OTFS-NOMA.

\section{Numerical Results and Discussion}

This section presents numerical results to showcase the effectiveness of the proposed OTFS-NOMA equalization and detection algorithm. As a benchmark, an OTFS-NOMA system using MMSE equalization and SIC for detection is considered, which is referred to as MMSE-SIC. Monte Carlo simulation is used to average the results over $10^5$ random channel realizations. A carrier frequency of $f_c = 5.9$ GHz, a transmission bandwidth of 4.95~MHz and a delay-Doppler grid size of $M=64$ and $N=16$ are considered. For power allocation, we use the average-SNR-based fractional transmit power allocation (FTPA) scheme outlined in \cite{Chatt_OTFS_NOMA}. Additionally, we consider a fixed SNR difference of 15~dB between the users, i.e., User 2 has an average SNR that is 15~dB higher than that of User 1. For each user, we consider a starting threshold of $T_i^{(1)} = 2d_i$. The Tapped Delay Line C (TDL-C) model with a delay spread of 300~ns \cite{3gpp_TS38901} is used for the channel model. We consider a range of maximum Doppler shifts from 500~Hz to 2500~Hz, which corresponds to velocities of approximately 90~km/h to 450~km/h at a carrier frequency of 5.9 GHz. The Doppler shifts are generated using Jakes' model \cite{jakes_model}. For the LSQR algorithm, a maximum number of iterations of $U=15$ and a tolerance of $\epsilon = 10^{-2}$ are considered, which are commonly used values for LSQR in related literature \cite{Qu_OTFS_detection, QU_ScFDMA}.   For Algorithm 1, we consider a maximum number of iterations of $K=10$ to limit the computational complexity. 
\begin{figure}[t]
    \centering
    \includegraphics[width = \columnwidth]{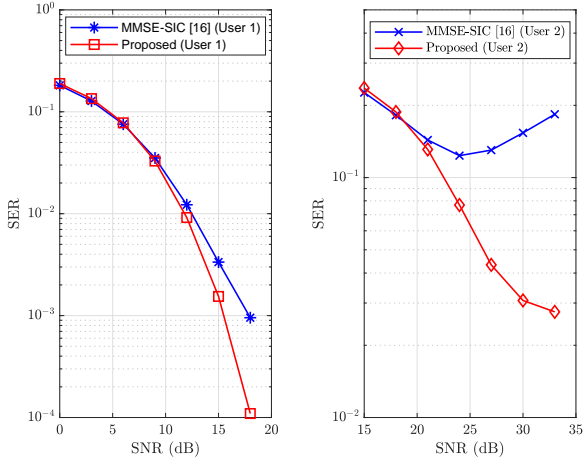}
    \caption{Comparison of the SER performance of the proposed detector with that of a detector using MMSE equalization and SIC, with different SNR levels, for the case where each user is allocated a 4-QAM constellation.}
    \label{fig:4QAM_SNR}
\end{figure}

\begin{figure}[t]
    \centering
    \includegraphics[width = \columnwidth]{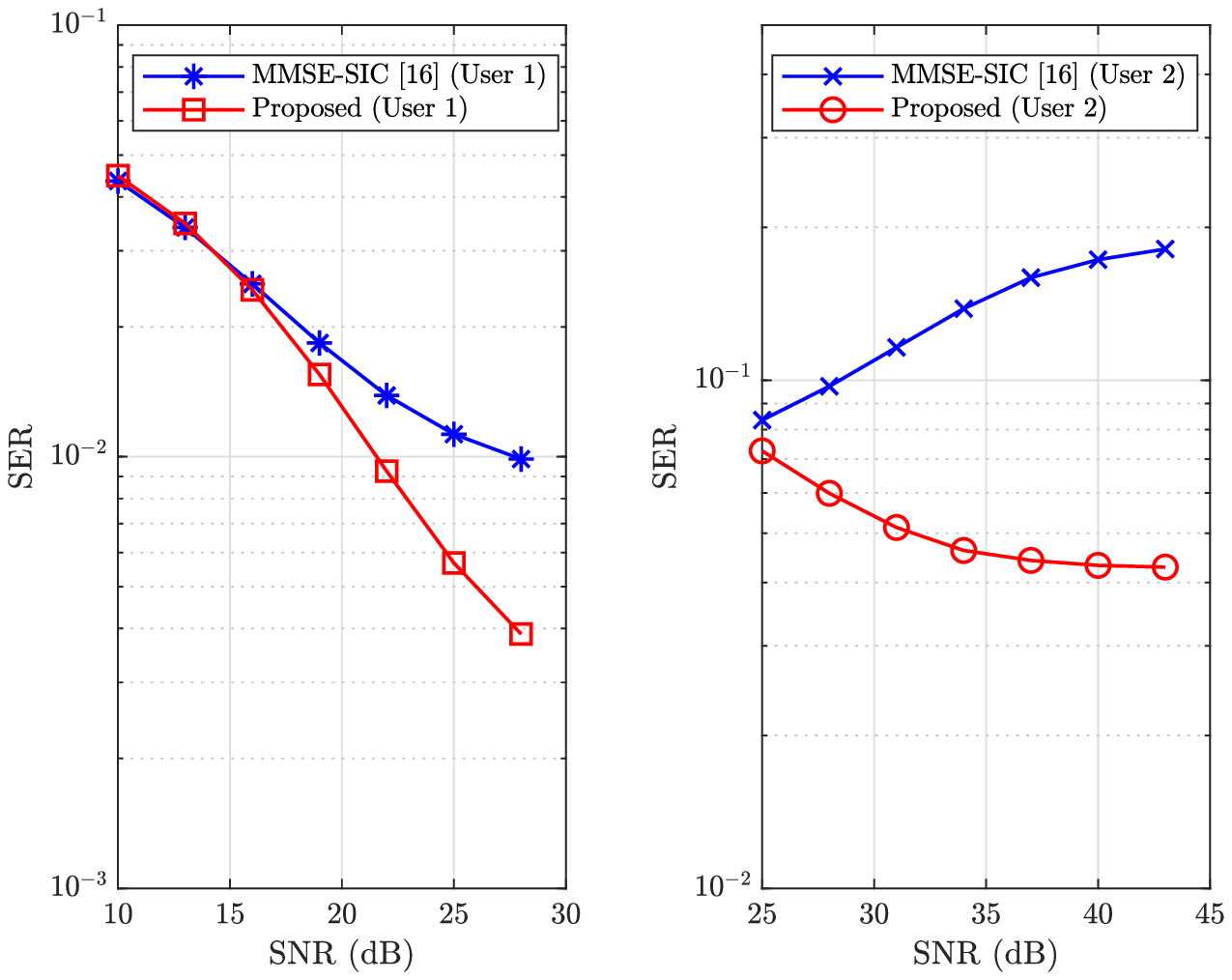}
    \caption{Comparison of the SER performance of the proposed detector with that of a detector using MMSE equalization and SIC, with different maximum Doppler shifts, for the case where each user is allocated a 16-QAM constellation.}
    \label{fig:16QAM_SNR}
\end{figure}

\begin{figure}[t]
    \centering
    \includegraphics[width = \columnwidth]{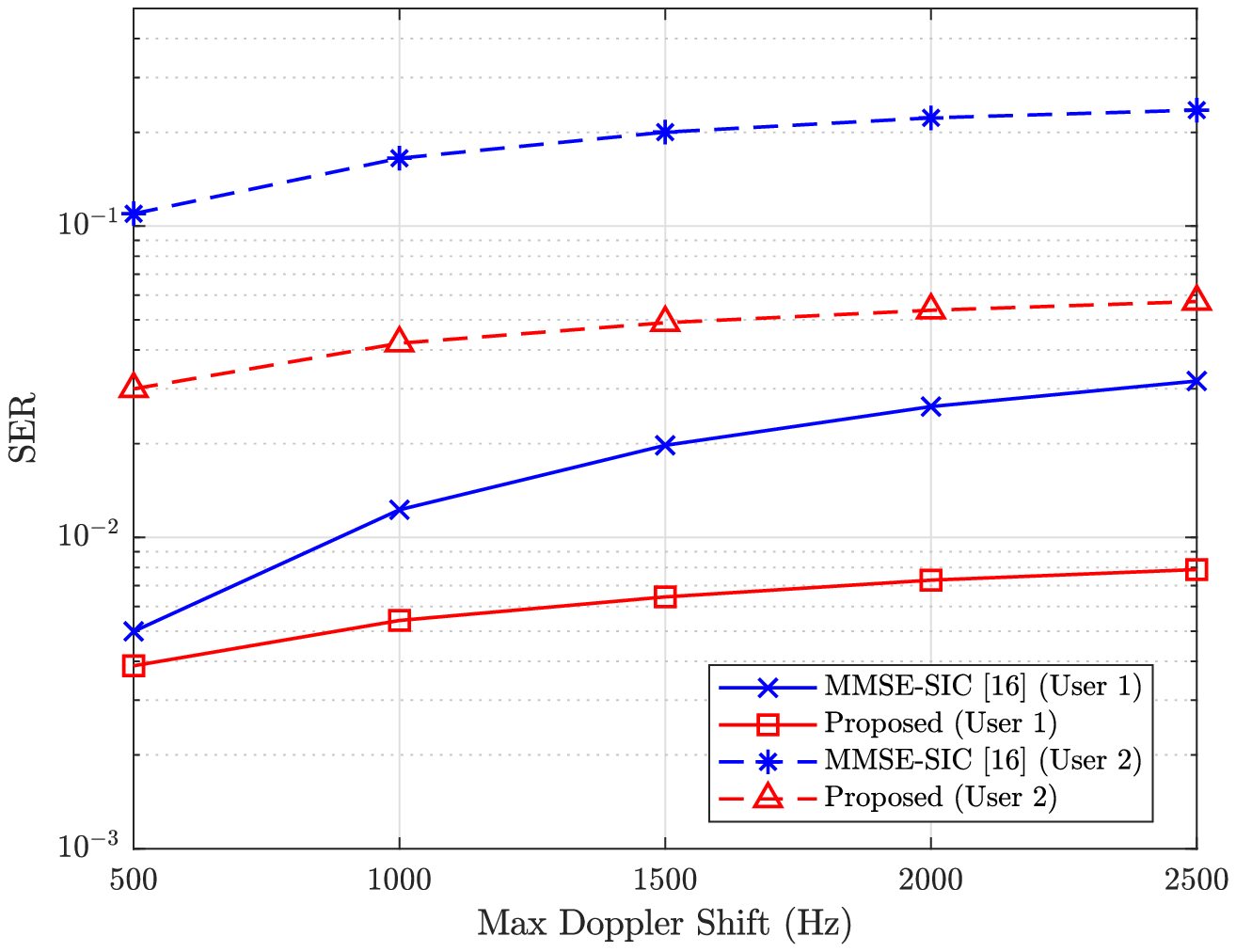}
    \caption{Comparison of the SER performance of the proposed detector with that of a detector using MMSE equalization and SIC, with different maximum Doppler shifts, for the case where each user is allocated a 16-QAM constellation.}
    \label{fig:16QAM_dopp}
\end{figure}

\begin{figure}[t]
    \centering
    \includegraphics[width = \columnwidth]{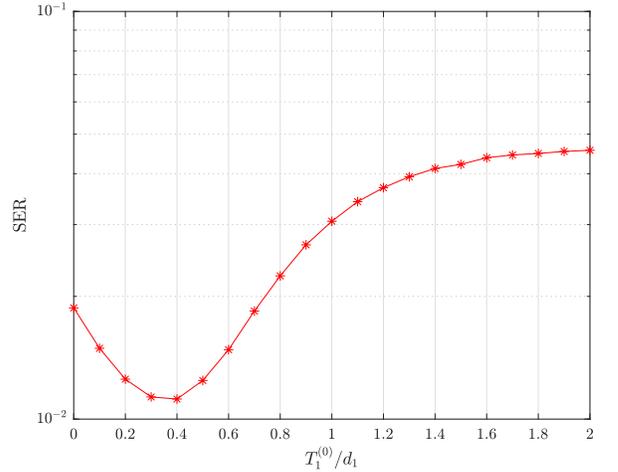}
    \caption{SER for User 2 using Algorithm 1 for different values of the User 1 starting threshold.}
    \label{fig:16QAM_thresh}
\end{figure}

Fig. \ref{fig:4QAM_SNR} shows the SER of both users under the proposed equalization and detection method, compared to the benchmark MMSE-SIC scheme, for different signal-to-noise ratio (SNR) conditions.  For these simulations each user's symbols are taken from a 4-QAM constellation, i.e., $\mathcal{A}_1 = \mathcal{A}_2 = 4$ and the user velocity is fixed at 200 km/h, which equates to a maximum Doppler shift of approximately 1000~Hz. It can be seen from Fig. \ref{fig:4QAM_SNR} that for both users, the proposed method outperforms the MMSE-SIC method, especially at high SNR. For User 1, the proposed method provides an SNR gain of 3~dB at an SER of $10^{-3}$ over MMSE-SIC. For User 2, the proposed method provides significant performance gains as MMSE-SIC performs very poorly, especially at high SNRs. This is due to the fact that, as the SNR increases, MMSE equalization becomes closer to zero-forcing equalization and the interference is amplified by the matrix inversion.

Fig. \ref{fig:16QAM_SNR} shows the SER of both users under the proposed equalization and detection method, compared to the benchmark MMSE-SIC scheme, for different SNR conditions, for the case where each user's symbols are taken from a 16-QAM constellation. As with the 4-QAM results, the user velocity is fixed at 200 km/h. It can be seen from Fig. \ref{fig:16QAM_SNR} that for both users the proposed method outperforms the MMSE-SIC method also for the 16-QAM scenario. For User 1, the proposed method provides an SNR gain of 6~dB at an SER of $10^{-2}$ over MMSE-SIC. For User 2, the proposed method provides significant performance gains over MMSE-SIC, which once again performs poorly due to the noise amplification caused by the matrix inversion.

Fig. \ref{fig:16QAM_dopp} shows the SER of both users under the proposed equalization and detection method, compared to the benchmark MMSE-SIC scheme, for different values of maximum Doppler shift, for the case where each user's symbols are taken from a 16-QAM constellation. Fixed SNRs of 20~dB and 35~dB are considered for User 1 and 2, respectively. It can be seen that the performance gains of the proposed method over MMSE-SIC actually improves in high Doppler environments, as the performance of MMSE-SIC deteriorates significantly at higher maximum Doppler shifts. This is because as the Doppler spread increases, the channel matrix is more likely to be ill-conditioned \cite{Qu_OTFS_detection}; hence, the matrix inversion involved in MMSE equalization may not be robust and can introduce significant equalization error. 

The performance of the proposed algorithm is dependent on the effectiveness of the interference cancellation process which is in turn dependent on the thresholds of the RZ detector. With this in mind, we have studied the effect of the choice of starting threshold on the performance of our proposed method. To demonstrate this, Fig. \ref{fig:16QAM_thresh} shows the SER of User 2 as a function of the starting User 1 threshold $T_1^{(1)}$, normalized by $d_1$. The results in this figure are for a fixed SNR of 35~dB and a velocity of 200~km/h. It can be seen from Fig. \ref{fig:16QAM_thresh} that the User 1 threshold has an effect on the detection performance of User 2 detection, and that, somewhat counter-intuitively, a tighter (conservative) User 1 threshold degrades the performance of User 2. This is because using tight starting thresholds means that fewer User 1 symbols are detected during early iterations and their MUI is still present in the system when the User 2 symbols are being detected. This highlights that the overall detection performance for both users' symbols is sensitive to the choice of the initial RZ thresholds, and that these should be chosen carefully to obtain the best performance.  

\section{Conclusion}
This paper has presented a novel low-complexity receiver for downlink OTFS-NOMA. The proposed method uses an iterative process which deploys the LSQR algorithm to equalize the channel, RZ detection to detect symbols from all users within the iterations and interference cancellation to remove MUI as well as IDI and ISI. Numerical results demonstrate the superiority of the proposed method, in terms of SER performance, with respect to an MMSE-SIC benchmark scheme. Furthermore, it has been demonstrated that the performance of the proposed algorithm is sensitive to the choice of starting thresholds, which need to be designed carefully.

\bibliographystyle{IEEEtran}
\bibliography{biblio}

\end{document}